\documentclass[9pt,twocolumn,twoside]{pnas-new}

\templatetype{pnasresearcharticle} 

\usepackage{longtable}

\begin{document}

\title{Two Distinct Populations of Dark Comets Delineated by Orbits and Sizes}

\author[a,b,c]{Darryl Z. Seligman}
\author[d]{Davide Farnocchia}
\author[e]{Marco Micheli}
\author[f]{Olivier R. Hainaut}
\author[g]{Henry H. Hsieh}
\author[a,h,i]{Adina D. Feinstein}
\author[d]{Steven R. Chesley}
\author[j]{Aster G. Taylor}
\author[l]{Joseph Masiero}
\author[m]{Karen J. Meech}

\affil[a]{Department of Physics and Astronomy, Michigan State University, East Lansing, MI 48824, USA;}
\affil[b]{Department of Astronomy and Carl Sagan Institute, Cornell University, 122 Sciences Drive, Ithaca, NY, 14853, USA;}
\affil[c]{NSF Astronomy and Astrophysics Postdoctoral Fellow}
\affil[d]{Jet Propulsion Laboratory, California Institute of Technology,
4800 Oak Grove Dr.,
Pasadena, CA 91109, USA}
\affil[e]{ESA NEO Coordination Centre, Largo Galileo Galilei 1,
 I-00044 Frascati (RM), Italy}
\affil[f]{European Southern Observatory, Karl-Schwarzschild-Strasse 2, Garching bei München, D-85748, Germany}
\affil[g]{Planetary Science Institute, 1700 East Fort Lowell Rd., Suite 106, Tucson, AZ 85719, USA}
\affil[h]{Laboratory for Atmospheric and Space Physics, University of Colorado Boulder, UCB 600, Boulder, CO 80309}
\affil[i]{NHFP Sagan Fellow}
\affil[k]{Dept. of Astronomy, University of Michigan, 323 West Hall, 1085 S University Ave, Ann Arbor, MI, 48109, USA; ORCiD: 0000-0002-0140-4475}
\affil[l]{Caltech/IPAC, 1200 E California Blvd, MC 100-22, Pasadena, CA 91125, USA}
\affil[m]{Institute for Astronomy, University of Hawaii, 2680 Woodlawn Dr., Honolulu, HI 96822, USA}

\leadauthor{Seligman}


\significancestatement{ Dark comets are near-Earth objects with no detected coma that have significant nongravitational accelerations explainable by outgassing of volatiles, analogous to the first interstellar object 1I/`Oumuamua. These objects represent a new --- and potentially widespread --- class of small bodies that further populate the continuum between asteroids and comets, and for which the active nature is inferred from their orbital motion. We report detections of seven new dark comets which demonstrate that there are two distinct populations based on their orbits and sizes. These objects represent a new class of Solar System objects that may have delivered material to the Earth necessary for the development of life such as volatiles and organics.}

\authordeclaration{The authors declare that they have no
competing financial interests.}
\equalauthors{\textsuperscript{1}D.~Z.~S. and D.~F. contributed equally to this work.  }
\correspondingauthor{\textsuperscript{2}To whom correspondence should be addressed. E-mail: dzs\@msu.edu}

\keywords{Asteroids $|$ Comets $|$ Interstellar Objects  }

\begin{abstract}
Small bodies are capable of delivering essential prerequisites for the development of life, such as volatiles and organics, to the terrestrial planets. For example, empirical evidence suggests that water was delivered to the Earth by hydrated planetesimals from distant regions of the Solar System \cite{Meech2020}.  Recently, several morphologically inactive near-Earth objects (NEOs) were reported to experience significant nongravitational accelerations inconsistent with radiation-based effects, and possibly explained by volatile-driven outgassing \cite{Farnocchia2022,Seligman2022}. However, these ``dark comets'' display no evidence of comae in archival images, which are the defining feature of cometary activity. Here we report detections of nongravitational accelerations on seven additional objects previously classified as inactive (doubling the population) that could also be explainable by asymmetric mass loss. A detailed search of archival survey and targeted data rendered no  detection of dust activity in any of these objects in individual or stacked images.  We calculate dust production limits of $\sim10$, $0.1$, and $0.1$~kg~s$^{-1}$  for  1998~FR$_{11}$, 2001~ME$_{1}$, and 2003~RM  with these data, indicating  little or no dust surrounding the objects during the observations. This set of dark comets  reveals the delineation between two distinct populations: larger, ``outer'' dark comets on eccentric orbits that are end members of a continuum in activity level of comets, and  smaller, ``inner'' dark comets on  near-circular orbits that could signify a new population.  These objects may trace  various stages in the life cycle of a previously undetected, but potentially numerous, volatile-rich population that may have provided essential material to the Earth. 
\end{abstract}

\dates{This manuscript was compiled on \today}
\doi{\url{www.pnas.org/cgi/doi/10.1073/pnas.2406424121}}

\maketitle
\thispagestyle{firststyle}
\ifthenelse{\boolean{shortarticle}}{\ifthenelse{\boolean{singlecolumn}}{\abscontentformatted}{\abscontent}}{}


\dropcap{N}ongravitational accelerations have previously been detected on  inactive asteroids. These accelerations   are  caused by either the Yarkovsky effect \cite{Vokrouhlicky2015_ast4} --- a predominantly transverse acceleration caused by anisotropic reradiation of thermal photons --- or radiation pressure \cite{Vokrouhlicky2000}. Nongravitational accelerations have been detected on over 200 near-Earth objects (NEOs) from the Yarkovsky effect \cite{Farnocchia2013,Greenberg2020} and on a handful of asteroids from radiation pressure \cite{Micheli2012,Micheli2013,Micheli2014,Mommert2014bd,Mommert2014md,Farnocchia2017TC25,Fedorets2020}.

Recently, seven apparently inactive NEOs were reported to exhibit statistically significant nongravitational accelerations \cite{Farnocchia2022,Seligman2022}. These accelerations  are not compatible with the radiation-driven forces that typically affect the motion of asteroids. These objects have been referred to  as ``dark comets" --- where comets is used as a synonym for volatile-rich bodies throughout this manuscript --- because their nongravitational accelerations are consistent with cometary outgassing with no observed associated dust production. The lack of observational confirmation of activity associated with these objects is likely partially due to the relatively shallow images obtained by all-sky surveys that dominate the observational record for most small asteroids and a lack of sufficiently sensitive targeted observations at appropriate times (e.g., near perihelion).
Given the strength of their nongravitational accelerations it is surprising that no activity was reported, even from relatively shallow all-sky survey data.  Along this line of reasoning, it is alternatively possible that these objects exhibit activity when not observed, such as in the case of 2014 XK$_6$ \cite{Masiero2018}. 
The seven objects are 2003 RM, 1998 KY$_{26}$, 2005 VL$_1$, 2016 NJ$_{33}$, 2010 VL$_{65}$, 2006 RH$_{120}$, and 2010 RF$_{12}$.  It is possible that these dark comets are representative of a  population of objects that come close to Earth and could have delivered volatile material.

The nongravitational accelerations and lack of any reports of activity in these dark comets are reminiscent of the first macroscopic interstellar object discovered traversing the inner solar system, 1I/`Oumuamua \cite{Williams17}. 1I/`Oumuamua also exhibited nongravitational acceleration inconsistent with radiation-based effects, albeit orders of magnitude larger than those measured in dark comets \cite{Micheli2018}. Deep images of 1I/`Oumuamua also displayed no photometric or morphological evidence for activity or dust production \cite{Meech2017,Ye2017,Jewitt2017}.  The magnitudes and directions of the nongravitational accelerations in the dark comets and in 1I/`Oumuamua are inconsistent with either radiation pressure or the Yarkovsky effect for typical cometary/asteroidal bulk material properties \cite{Micheli2018,Farnocchia2022,Seligman2022}. Therefore, it is likely that the nongravitational accelerations of dark comets are due to outgassing, despite their lack of observed dust production typically associated with volatile outgassing.

The first  identified dark comet was 2003 RM \cite{Farnocchia2022}. This object has a radius $r_{n}\sim150$ m  and exhibits highly significant (60-$\sigma$) nongravitational acceleration in the orbital transverse direction \cite{Farnocchia2022}. The remaining previously reported dark comets are smaller and predominantly exhibit accelerations out of their orbital plane only. These out-of-plane accelerations could be caused by seasonally-induced polar outgassing for rapidly rotating objects \cite{Taylor2023}. In this paper, we report new detections of nongravitational accelerations on seven additional NEOs. 
\begin{figure*}
\centering
\includegraphics[width=1.0\linewidth]{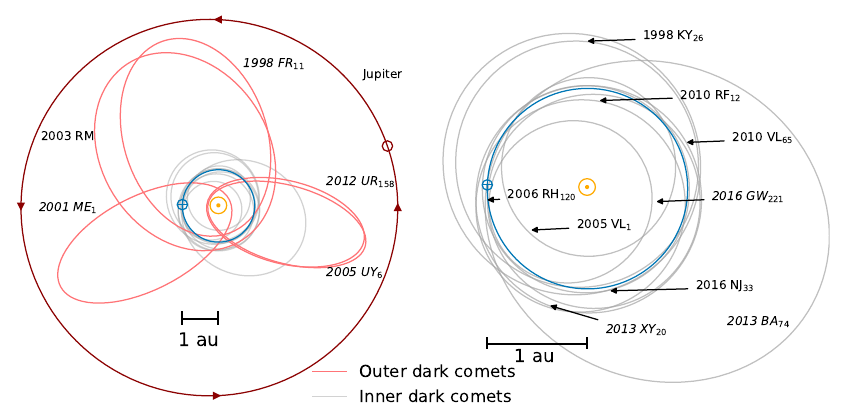}
\caption{The orbits of the fourteen currently known  dark comets, Jupiter, and the Earth.  One orbital period is shown for each object starting on the vernal equinox in 2023 March, and the points represent  the corresponding initial positions of the Earth and of Jupiter. The left and right panels highlight the orbits of the outer and inner dark comets respectively.  Orbits of the newly discovered dark comets are labeled with italicized font, and the orbits of previously reported objects are labeled in roman font. }
\label{fig:orbits}
\end{figure*}

\setlength{\tabcolsep}{1.5pt}
\begin{table}
\begin{center}   
\caption{The measured orbital and physical properties of the dark comets. The columns correspond to the name, semimajor axis, eccentricity, inclination, perihelion, Tisserand parameter with respect to Jupiter, H magnitude in V-band, rotational period, albedo, nuclear radius, and percentage chance of originating as a Jupiter Family Comet (JFC). Albedos and radii are from \cite{Ostro1999,Masiero2020,Masiero2021}. (i) Outer, (ii) inner, and (iii) newly discovered dark comets are indicated with (i) bold,  (ii) regular, and (iii) italicized font in all tables.}
\label{table:physical_properties}
\footnotesize
\begin{tabular}{cccccccccccccc l@{}c@{}lc l@{}c@{}lc l@{}c@{}lc c c} 
name&a&e&i&q&T$_J$&H&P$_{\rm Rot}$&albedo&r$_n$&P$_{\rm JFC}$& \\ 
&[au]&&[deg]&[au]&&[Mag]&[h]&&[km]&[\%]& \\ 
\textit{\textbf{2001 ME$_{1}$}}&2.65&0.87&5.80&0.36&2.67&16.53&&0.018&5$\pm$2&11.5& \\ 
\textit{\textbf{2005 UY$_{6}$}}&2.26&0.87&12.15&0.29&2.94&18.14&&&&0& \\ 
\textit{\textbf{1998 FR$_{11}$}}&2.81&0.71&6.66&0.83&2.88&16.42&&0.02&5$\pm$2&2.6& \\ 
\textit{\textbf{2012 UR$_{158}$}}&2.24&0.86&3.22&0.32&3.00&20.7&&&0.63&0& \\ 
\textbf{2003 RM}&2.92&0.60&10.85&1.17&2.96&19.8&&&0.23&3.9& \\ 
2005 VL$_{1}$&0.89&0.23&0.25&0.69&6.64&26.45&&&&0& \\ 
2010 RF$_{12}$&1.06&0.19&0.88&0.86&5.79&28.42&&&&0& \\ 
1998 KY$_{26}$&1.23&0.20&1.48&0.98&5.18&25.6&0.178&&0.015&0& \\ 
2016 NJ$_{33}$&1.31&0.21&6.64&1.04&4.95&25.49&0.41-1.99&&&0& \\ 
2010 VL$_{65}$&1.07&0.14&4.41&0.91&5.76&29.22&&&&0& \\ 
\textit{2013 BA$_{74}$}&1.75&0.44&5.30&0.98&4.00&25.4&&&&0& \\ 
2006 RH$_{120}$&1.00&0.04&0.31&0.96&6.09&29.5&0.046&&&0& \\ 
\textit{2016 GW$_{221}$}&0.83&0.27&3.65&0.61&7.05&24.76&0.2856&&&0& \\ 
\textit{2013 XY$_{20}$}&1.13&0.11&2.86&1.01&5.52&25.65&&&&0& \\  
\hline
\end{tabular}
\end{center}
\end{table}

\setlength{\tabcolsep}{1.5pt}
\begin{table*}
\begin{center}   
\caption{The derived nongravitational accelerations normalized at 1 au of each dark comet. The table shows the three components' magnitude, uncertainty and associated significance ($\sigma$), which is the ratio of the uncertainty to the magnitude. The weighted $\chi^2$ value of the astrometric fitting procedure with both the gravity only case ($\chi^2_{G}$) and the nongravitational acceleration included case ($\chi^2_{NG}$) are shown in the last two columns.} 
\label{table:accelerations}
\footnotesize
\begin{tabular}{lccrccrccrcccc l@{}c@{}lc l@{}c@{}lc l@{}c@{}lc c c} 
name&A$_1$&Error&$\sigma$&A$_2$&Error&$\sigma$&A$_3$&Error&$\sigma$&$\chi^2_{G}$&$\chi^2_{NG}$   \\ 
&[au d$^{-2}$]&[au d$^{-2}$]&&[au d$^{-2}$]&[au d$^{-2}$]&&[au d$^{-2}$]&[au d$^{-2}$]&& \\ 
\textit{\textbf{2001 ME$_{1}$}}&$-2.47 \times 10^{-13}$&$\phantom{-}1.37 \times 10^{-10}$&$0.0$&$-2.54 \times 10^{-13}$&$\phantom{-}2.27 \times 10^{-14}$&$11.18$&$\phantom{-}2.82 \times 10^{-12}$&$\phantom{-}1.78 \times 10^{-12}$&$1.58$&355&204& \\ 
\textit{\textbf{2005 UY$_{6}$}}&$-2.92 \times 10^{-10}$&$\phantom{-}4.28 \times 10^{-10}$&$0.68$&$-6.35 \times 10^{-13}$&$\phantom{-}1.59 \times 10^{-13}$&$4.0$&$-4.86 \times 10^{-12}$&$\phantom{-}6.89 \times 10^{-12}$&$0.71$&91&65 \\ 
\textit{\textbf{1998 FR$_{11}$}}&$-1.83 \times 10^{-10}$&$\phantom{-}2.96 \times 10^{-10}$&$0.62$&$\phantom{-}2.94 \times 10^{-13}$&$\phantom{-}9.56 \times 10^{-14}$&$3.08$&$\phantom{-}8.38 \times 10^{-12}$&$\phantom{-}8.62 \times 10^{-12}$&$0.97$&144&130 \\ 
\textit{\textbf{2012 UR$_{158}$}}&$\phantom{-}8.18 \times 10^{-11}$&$\phantom{-}6.66 \times 10^{-10}$&$0.12$&$-7.59 \times 10^{-12}$&$\phantom{-}4.72 \times 10^{-13}$&$16.07$&$\phantom{-}7.15 \times 10^{-12}$&$\phantom{-}3.74 \times 10^{-12}$&$1.91$&1687&38 \\ 
\textbf{2003 RM}&$-3.96 \times 10^{-11}$&$\phantom{-}1.10 \times 10^{-10}$&$0.36$&$\phantom{-}2.10 \times 10^{-12}$&$\phantom{-}2.80 \times 10^{-14}$&$75.03$&$\phantom{-}5.33 \times 10^{-13}$&$\phantom{-}4.10 \times 10^{-12}$&$0.13$&12265&162 \\ 
2005 VL$_{1}$&$-8.30 \times 10^{-10}$&$\phantom{-}7.59 \times 10^{-10}$&$1.09$&$-8.32 \times 10^{-13}$&$\phantom{-}5.61 \times 10^{-13}$&$1.48$&$-2.41 \times 10^{-11}$&$\phantom{-}3.95 \times 10^{-12}$&$6.12$&219&29 \\ 
2010 RF$_{12}$&$\phantom{-}3.40 \times 10^{-11}$&$\phantom{-}5.82 \times 10^{-11}$&$0.58$&$-2.12 \times 10^{-13}$&$\phantom{-}2.12 \times 10^{-13}$&$1.0$&$-1.51 \times 10^{-11}$&$\phantom{-}1.42 \times 10^{-12}$&$10.65$&173&60 \\ 
1998 KY$_{26}$&$\phantom{-}1.60 \times 10^{-10}$&$\phantom{-}8.77 \times 10^{-11}$&$1.83$&$-1.38 \times 10^{-13}$&$\phantom{-}5.67 \times 10^{-14}$&$2.43$&$\phantom{-}2.70 \times 10^{-11}$&$\phantom{-}6.45 \times 10^{-12}$&$4.19$&89&33 \\ 
2016 NJ$_{33}$&$\phantom{-}9.48 \times 10^{-10}$&$\phantom{-}2.93 \times 10^{-10}$&$3.24$&$-5.49 \times 10^{-13}$&$\phantom{-}1.91 \times 10^{-13}$&$2.87$&$\phantom{-}8.49 \times 10^{-11}$&$\phantom{-}1.63 \times 10^{-11}$&$5.21$& 132&53\\ 
2010 VL$_{65}$&$\phantom{-}6.59 \times 10^{-10}$&$\phantom{-}1.30 \times 10^{-9}$&$0.51$&$\phantom{-}1.45 \times 10^{-13}$&$\phantom{-}5.34 \times 10^{-13}$&$0.27$&$-9.12 \times 10^{-11}$&$\phantom{-}1.30 \times 10^{-11}$&$7.02$& 4441&28\\ 
\textit{2013 BA$_{74}$}&$\phantom{-}2.47 \times 10^{-10}$&$\phantom{-}1.84 \times 10^{-9}$&$0.13$&$\phantom{-}3.66 \times 10^{-13}$&$\phantom{-}4.57 \times 10^{-13}$&$0.8$&$\phantom{-}1.92 \times 10^{-11}$&$\phantom{-}5.30 \times 10^{-12}$&$3.61$&63&25 \\ 
2006 RH$_{120}$&$\phantom{-}1.38 \times 10^{-10}$&$\phantom{-}7.79 \times 10^{-12}$&$17.73$&$-5.07 \times 10^{-11}$&$\phantom{-}6.37 \times 10^{-12}$&$7.96$&$-1.30 \times 10^{-11}$&$\phantom{-}3.32 \times 10^{-12}$&$3.92$& 1579&65\\ 
\textit{2016 GW$_{221}$}&$\phantom{-}8.73 \times 10^{-11}$&$\phantom{-}8.81 \times 10^{-11}$&$0.99$&$-1.85 \times 10^{-13}$&$\phantom{-}2.45 \times 10^{-14}$&$7.53$&$\phantom{-}3.50 \times 10^{-11}$&$\phantom{-}8.03 \times 10^{-12}$&$4.36$& 220&60\\ 
\textit{2013 XY$_{20}$}&$\phantom{-}1.81 \times 10^{-10}$&$\phantom{-}4.98 \times 10^{-11}$&$3.64$&$-1.52 \times 10^{-12}$&$\phantom{-}2.75 \times 10^{-13}$&$5.55$&$-1.09 \times 10^{-10}$&$\phantom{-}1.29 \times 10^{-11}$&$8.43$& 129&63\\ 
\hline
\end{tabular}
\end{center}
\end{table*}

\section*{Results}
\subsection*{Delineation of the Two Populations Based on their Orbits and Sizes}

Using the techniques described in the Methods section, we determined nongravitational accelerations for seven additional NEOs using all astrometric data published by the Minor Planet Center (MPC)\cite{Ostro2002} and radar measurements\footnote{\url{https://ui.adsabs.harvard.edu/abs/2002aste.book..151O/abstract}}. In Table \ref{table:physical_properties} we list  the new  in comparison with the known dark comets and their physical properties. In Table \ref{table:accelerations} we report our new measurements of the three components of the nongravitational accelerations  of these objects. The three directions of the accelerations --- A$_1$, A$_2$ and A$_3$ --- correspond to the  radial, transverse, and out-of-plane directions (see Methods section). In Figure \ref{fig:orbits} we show the orbits of all of the currently known dark comets, as well as that of the Earth and of Jupiter.

\begin{figure}
\centering
\includegraphics[width=1.\linewidth]{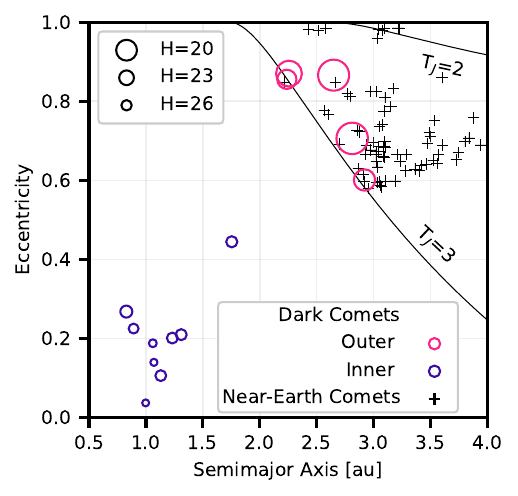}
\caption{Location of all dark comets and near-Earth comets in semimajor axis and eccentricity space.  The region where the Tisserand parameter with respect to Jupiter (T$_J$, Equation \ref{eq:tisserand}) satisfies $2<T_J<3$ is the classical definition of a JFC. The inner dark comets are smaller (larger H mag) than the outer dark comets and unlike the outer dark comets cannot be sourced from the JFC population.}
\label{fig:tisserand}
\end{figure}

The total sample of  fourteen dark comets appears to fall into two distinct populations: outer dark comets --- large objects with Jupiter Family Comet (JFC) orbits ---  and inner dark comets --- smaller objects on orbits with lower eccentricity and smaller semimajor axis. The left and right panels of Figure \ref{fig:orbits} are zoomed-out and -in in order to highlight both distinct populations. Notably, the orbits and sizes of outer dark comets resemble 2003 RM while the inner dark comets largely resemble the other six previously identified objects.  This is summarized in Figure \ref{fig:tisserand}, which shows the location of all of the dark comets and near-Earth comets in semimajor axis and eccentricity space.

\begin{figure}
\centering
\includegraphics[width=1.\linewidth]{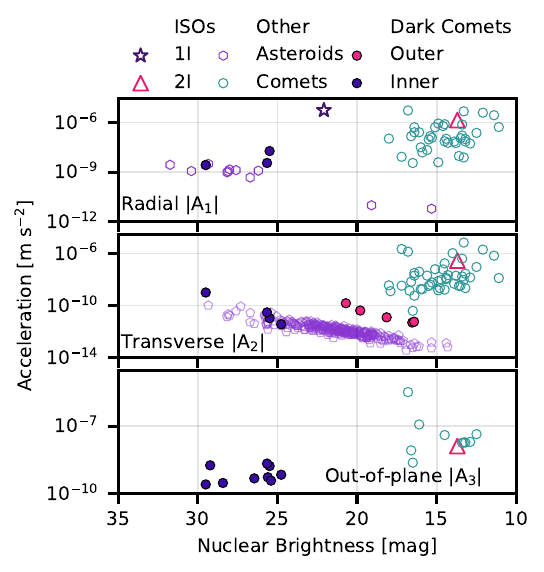}
\caption{Like other small bodies, dark comets show significant nongravitational accelerations. The nuclear brightness is represented by H mag for inactive objects and the $M_2$ comet nuclear magnitude parameter is shown for active objects. Only nongravitational accelerations with  $>3-\sigma$ significance are shown. Data for  comets and asteroids are  from the JPL Small Body Database.} 
\label{fig:3panel}
\end{figure}

The differences in the two populations is also apparent when considering their Tisserand parameter with respect to Jupiter \cite{tisserand1896_tj,Murray1999}, $T_J$, which is defined as
\begin{equation}\label{eq:tisserand}
    T_J=a_J/a+2\cos(i)\sqrt{a/a_j(1-e^2)}\,.
\end{equation}
 In Equation (\ref{eq:tisserand}), $a_J$ is Jupiter's semimajor axis, and $a$, $e$, and $i$ are the semimajor axis, eccentricity, and inclination of the orbit of the object. The Tisserand parameter with respect to a perturber (in this case Jupiter) is approximately conserved during a close interaction. JFCs that have been scattered into the inner solar system are typically defined by $2<T_J<3$  \cite{Kresak1972,Vaghi1973,Levison1996}, as shown in Figure  \ref{fig:tisserand}. It is evident that the outer dark comets, along with the near-Earth comets,  approximately reside within this region while the inner dark comets do not. We report the probability that each object originated in the JFC environment in Table \ref{table:physical_properties} (see Methods section). The only objects with nonzero chance of originating as JFCs are the outer dark comets. 

 \begin{figure*}
\centering
\includegraphics[width=1.\linewidth]{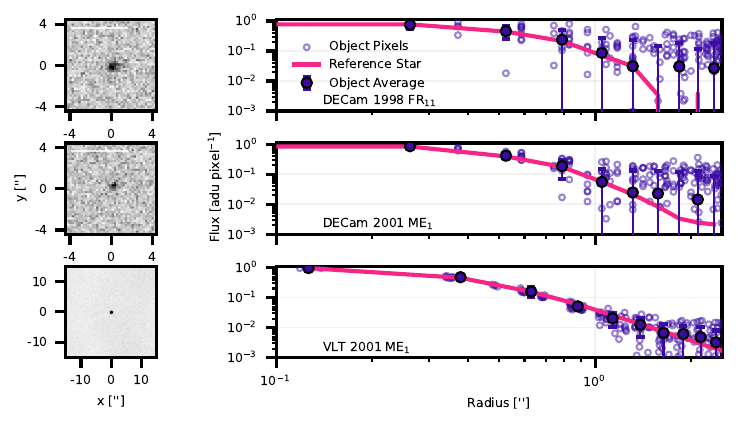}
\caption{Images of two dark comets from archival DECam data and our VLT data show no obvious evidence of faint coma either in the images (left) or in the surface brightness profiles compared to field stars (right).  The DECam images of 2001 ME$_1$ and 1998 FR$_{11}$ are from 2022-May-12 08:22 and 2016-Jan-15 05:55 when the objects were at  heliocentric distances of $\sim3.5$ and $\sim4.1$ au. The VLT image of 2001~ME$_1$ is a stack of images totalling 1500s of total exposure time using the FOcal Reducer and low dispersion Spectrograph (FORS2) on the ESO Very Large Telescope at Paranal.}
\label{fig:images_and_profiles}
\end{figure*}

There is a tentative distinction in the two populations' absolute magnitudes (which corresponds to size, assuming uniform albedo using Equation 2 in \cite{Pravec2007}) and possibly direction of the nongravitational accelerations (see Figure \ref{fig:3panel}). The outer dark comets have H magnitudes between $\sim16-21$, while the inner dark comets have H magnitudes between $\sim25-30$. However, this is likely partly due to an observational selection effect since it is more difficult to detect nongravitational accelerations on distant, faint objects. The  lack of large objects in the inner population is likely a physical effect because these would be well within observational limits. The outer dark comets all exhibit transverse nongravitational accelerations,  while the inner dark comets exhibit radial, transverse and out-of-plane accelerations.

\subsection*{Search for Activity}
 In an effort to investigate whether past activity associated with these objects has been missed, we performed a search for activity associated with these objects in archival data, specifically observations obtained by the Dark Energy Camera (DECam) on the 4 m Victor M.\ Blanco telescope in Chile and Megacam on the 3.6~m Canada-France-Hawaii Telescope in Hawaii at times when these objects were in those imagers' fields of view. We focused on these two instruments because they are large field of view imagers in regular use on large-aperture telescopes.  In Figure \ref{fig:images_and_profiles}, we show two example images  in which  the two brightest dark comets, 2001 ME$_{1}$ and 1998 FR$_{11}$, were recovered.  We also included a dataset on 2001~ME$_1$  obtained using the FOcal Reducer and low dispersion Spectrograph (FORS2) on the ESO Very Large Telescope at Paranal.

We constrain the dust production by comparing the photometric profile of an object  to that of background stars (see Methods). Example profile comparisons are shown in Figure~\ref{fig:images_and_profiles}. Although we only show the radial profile for two DECam images, all remaining images have radial profiles consistent with the stellar profiles. This  analysis incorporates all the information from the profile in a consolidated way, leading to an upper limit on the dust. Because the contribution of each radius is based on the noise level at that radius, it is not meaningful to report a single surface brightness limit. Therefore, in Table \ref{table:dustMassSummary} we show upper dust mass limits for each of these objects based on the nondetections of activity (see Methods).  In Table \ref{table:dustMassDetail}, we provide the details for every image analyzed in which one of these objects was recovered. The  upper limits on the dust production are calculated by dividing the mass limit by the exposure time and are of the order of $\sim10$, $0.1$ and $0.1$~kg~s$^{-1}$ at heliocentric distances of $\sim3.2-4.8$, $\sim4.1$ and $\sim1.2-1.6$ au  for  1998~FR$_{11}$, 2001~ME$_{1}$, and 2003~RM.

\begin{table}
\begin{center}   
\caption{Upper limit on the mass of dust that is compatible with the observations for 3 outer dark comets. The table reports the values obtained from the most constraining individual image, and from the combined profiles. "Inner profile" refers to the region within $1''$ from the center of the object, and "outer profile" to the ring between 1 and $2''$. Objects not listed in the table did not have images suitable for this analysis.} 
\label{table:dustMassSummary}
\footnotesize
\begin{tabular}{lcccrccccc} 
    Name      & \multicolumn{2}{c}{Best individual image}&  \multicolumn{3}{c}{Combined images}\\
    & M (inner) & M (outer)&tot.Exp.time & M (inner) & M (outer)   \\
    & [kg]& [kg]&  [s]&  [kg]& [kg] \\
  \textbf{\textit{1998 FR$_{11}$}} & 9.$\times 10^{3}$ & 1.$\times 10^{4}$ & 165  &  5.7$\times 10^{3}$&  1.9$\times 10^{3}$\\
  \textbf{\textit{2001 ME$_{1}$}}  & 2.1$\times 10^{3}$ & 9.5$\times 10^{2}$ & 2305  &  1.1$\times 10^3$&  2.1$\times 10^{2}$\\
\textbf{2003~RM} &  $7.9\times 10^1$& $3.8\times 10^1$& 142  &   $7.8\times 10^1$&  $2.5\times 10^1$\\
\end{tabular}
\end{center}
\end{table}

In order to perform a deeper search for activity using archival data, we use a novel enhanced approach to shifting and adding images that we refer to as `hyperstacking', where images from many different objects observed at many different times are combined to search for extremely faint activity. Hyperstacking exploits the tendency for extended cometary features (i.e., tails and trails) to align with the directions of either the projected antisolar or negative heliocentric velocity vectors on the sky (the latter of which also coincides with the object's orbit), or sometimes both \citep[e.g.,][]{hsieh2012_288p}.  
Since a solar system object's observing geometry typically does not vary significantly over the course of several hours or even days, it is generally unnecessary to consider expected tail or trail directions in standard stacking analyses that add images obtained over relatively short periods of time.  For images of many different objects obtained at many different times, however, the expected tail and trail directions could vary widely, meaning that simply stacking such images with no additional adjustments would spread any excess surface brightness flux from faint tails or trails over all azimuth angles.  By rotating images prior to stacking such that the expected tail or trail directions are aligned, hyperstacking seeks to concentrate excess surface brightness flux from faint tails or trails to a specific azimuth angle (in our case, 270$^{\circ}$ counterclockwise from the upward vertical direction, or directly to the right).

\begin{figure}
\centering
\includegraphics[width=1.\linewidth]{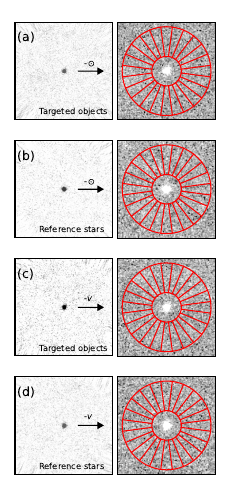}
\caption{Hyperstacked images combining $g'$-, $r'$-, $i'$-, and $z'$-band DECam images (200 pixels, or $50''$, on each side), collectively comprising 948~s of total exposure time, obtained between UT 2013 August 6 and UT 2022 May 12 of 1998 FR$_{11}$, 2001 ME$_{1}$, 2003 RM, and 2012 UR$_{158}$ with targets at the center of each panel. Panels (a) and (b) show data that have been rotated to align the antisolar vector ($-\odot$) as projected on the sky to the right in each individual image, and panels (c) and (d) show data that  have been rotated to align the negative heliocentric velocity vector ($-v$) as projected on the sky to the right in each individual image.  Panels (a) and (c) show stacks of images of target objects, and panels (b) and (d) show stacks of images of field stars (one per target detection) similar in brightness to our target objects from the same data that are provided for comparison.
Subpanels show (left) composite hyperstacked images and (right) diagrams of the portions of sky sampled for azimuthal sky surface brightness analysis in Figure \ref{fig:hyperstack_decam_profiles}.}
\label{fig:hyperstack_decam_images}
\end{figure}

\begin{figure}
\centering
\includegraphics[width=1.\linewidth]{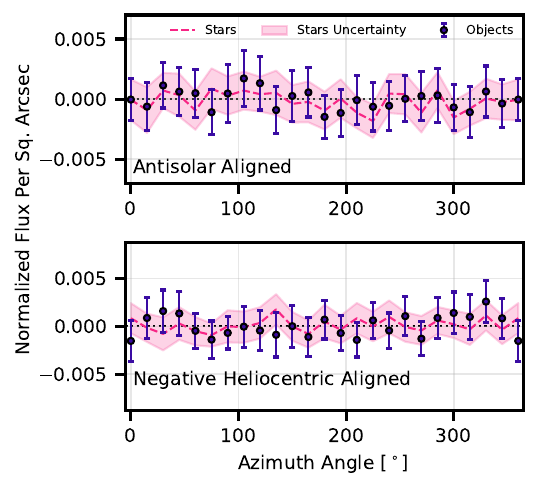}
\caption{The net surface brightness fluxes per square arcsecond normalized to the brightness of the central source as functions of azimuthal angle (in degrees counterclockwise from the upward vertical direction). The upper and lower panels correspond to the antisolar vector aligned (panels (a) and (b) in Figure \ref{fig:hyperstack_decam_images}) and negative heliocentric velocity vector aligned (panels (c) and (d) in Figure \ref{fig:hyperstack_decam_images}) cases. The profiles and associated uncertainties of the targeted dark comets (panels (a) and (c) in Figure \ref{fig:hyperstack_decam_images}) and stars (panels (b) and (d) in Figure \ref{fig:hyperstack_decam_images}) are shown in purple points and pink shaded regions respectively. }
\label{fig:hyperstack_decam_profiles}
\end{figure}

Following this approach (described in greater detail in Methods), we produced hyperstacks using data from DECam (Figures~\ref{fig:hyperstack_decam_images} and \ref{fig:hyperstack_decam_profiles}). No extended activity is immediately visible in the hyperstack, but to perform a more quantitative search for activity in the DECam hyperstack, we  performed measurements of the net surface brightness around the central stacked source as a function of azimuth angle.  We (i) divided the sky into wedges 15$^{\circ}$ in angular width and extending from $5''$ to $15''$ from the center of the image (see right subpanels in Figure~\ref{fig:hyperstack_decam_images}), (ii) measured the median flux per square arcsecond in each wedge, (iii) subtracted the median flux per square arcsecond of all wedges, (iv) normalized the remaining net flux to the brightness of the central stacked source within a $4''$ photometry aperture, and (v) plotted these normalized net fluxes per square arcsecond as functions of azimuth angle for hyperstacks of both our target objects and reference field stars aligned to their respective antisolar directions and negative heliocentric velocity directions (Figure~\ref{fig:hyperstack_decam_profiles}).

Dust emission aligned with the antisolar direction or the negative heliocentric velocity vector should appear as excess surface brightness at an azimuth angle of $\sim270^{\circ}$ in the purple points in Figure~\ref{fig:hyperstack_decam_profiles}.  No such excess brightness at that azimuth angle (or at any other azimuth angle) is detected within $1-\sigma$ uncertainties.  Azimuthal surface brightness plots for reference star stacks show similar azimuthal brightness uniformity in the surrounding sky, providing some assurance that the hyperstacking process did not introduce any notable systematics or artifacts into the final composite images. As the image levels were normalized to the flux of the object, the non-detection corresponds to $\sim$0.005 of that flux, but cannot be converted to a value in magnitude. 

It should be noted that the majority  of these observations were not conducted near perihelion where maximum activity detectability would be expected. While the  data we analyzed continues to support the classification of these objects as inactive, continued monitoring of these objects, especially near perihelion, is needed to more firmly establish their inactive statuses or to identify low levels of activity.

\section*{Discussion}

In accordance with previous results \cite{Seligman2022}, both the implied mass-loss rate, $dM_{\rm Nuc}/dt$, and  the timescale to lose all mass, $\tau_M$, of the dark comets are perplexing. The mass-loss rate may be estimated by equating the  outgassing force and the rate of nucleus momentum change using:
\begin{equation}\label{eq:outgassingrate}
\frac{d M_{\rm Nuc}}{dt}\simeq  \, M_{\rm Nuc}\,\bigg(\,\frac{|A|}{v_{\rm Gas} \zeta} \,\bigg)\,.
\end{equation}
In Equation \ref{eq:outgassingrate}, $M_{\rm Nuc}$ is the mass of the nucleus, $v_{\rm Gas}$ is the velocity of the  outgassing species, $|A|$ is the total magnitude of nongravitational acceleration calculated by summing all significant components in quadrature, and  $\zeta$ indicates the isotropy of the outflow, where  $\zeta$ = 1 corresponds to a  collimated outflow and $\zeta$ = 0.5 corresponds to an  isotropic hemispherical outflow.  
The characteristic mass-loss timescale $\tau_M$ is approximately given by, 
\begin{equation}\label{eq:timescale}
 \tau_M \simeq  M_{\rm Nuc} \bigg/ \,\bigg(\,\frac{d M_{\rm Nuc}}{dt} \,\bigg)\, \simeq  \,\bigg(\,\frac{v_{\rm Gas} \zeta}{|A|} \,\bigg)\,.
\end{equation}
 
\begin{figure}
\centering
\includegraphics[width=1.\linewidth]{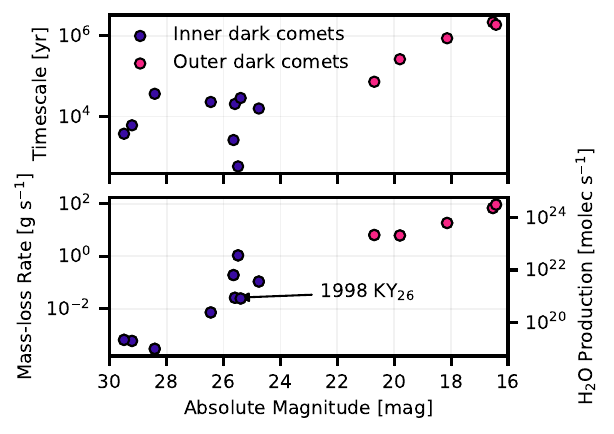}
\caption{The mass-loss rate and timescale for the dark comets (Equations \ref{eq:outgassingrate} and \ref{eq:timescale}) imply that they were either  (i)  not always losing mass at this rate or (ii)  only recently emplaced on these orbits.}  The inferred H$_2$O production rate shown on the right y-axis of the lower plot is calculated assuming that all mass-loss is due to H$_2$O sublimation. Hayabusa2$\#$ should readily detect this level of activity on 1998 KY$_{26}$. 
\label{fig:mass_loss_rate}
\end{figure}

Both of these quantities are shown for inner and outer dark comets in Figure \ref{fig:mass_loss_rate}, assuming $\zeta=1$ and  $v_{\rm Gas}=350$ m/s. To calculate the mass-loss rate, we first convert the H magnitude to a diameter assuming uniform  albedo of 0.1, and then use this value to estimate the nucleus mass assuming a bulk density of $\rho_{\rm Bulk} = 1\, {\rm g\, cm^{-3}}$. Even for the largest outer dark comets, the outgassing rates are orders of magnitude lower than the dust production limits calculated here implying upper limits of the dust to gas ratio $<10^4$, which is not constraining given typical cometary values of $\sim10^{-2}$. The mass-loss timescale ranges from $ \tau_M \sim10^4-10^5$ yr for inner dark comets to $ \tau_M \sim10^5-10^6$ yr for outer dark comets. As discussed in \cite{Seligman2022}, if the inner dark comets  are outgassing, then they  either (i) were not always outgassing at this rate or (ii) were only recently emplaced on these orbits.    

The search for activity in archival data presented in this paper is limited for several reasons, and our results showing no evidence of activity does not preclude the possibility that these objects are weakly active. First, the serendipitous apparitions of these objects in archival data is for the most part not coincident with their perihelia, where  any sublimation-driven activity is expected to be strongest.
Second, the sensitivity of activity searches with archival data is limited by the exposure times of those data, meaning that faint activity that could have been detected by longer exposures could be missed.
Finally, the lack of non-sidereal tracking for archival data also leads to trailing losses in terms of activity detectability. For these reasons, follow-up and targeted activity searches would be beneficial for any of these objects to more firmly characterize their active statuses.

\begin{figure*}
\centering
\includegraphics[width=1.\linewidth]{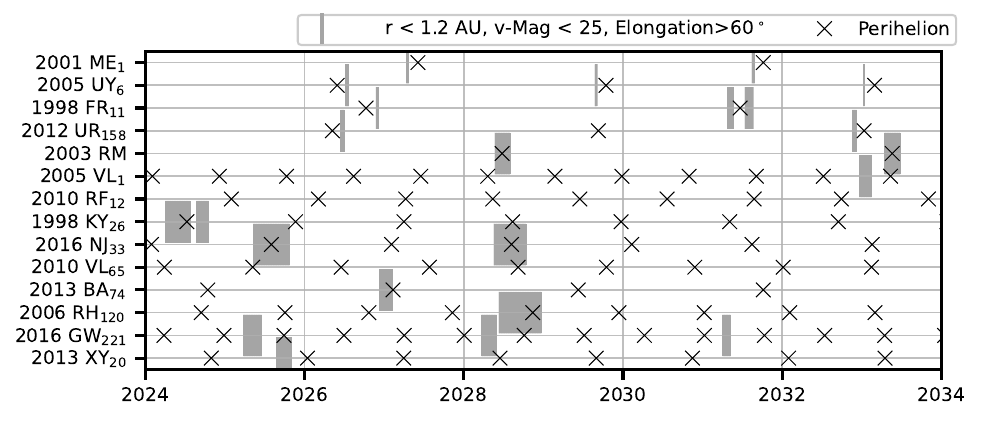}
\caption{Future observability of all of the dark comets  between 2024-2034. There will be narrow windows when meaningful observations can be made with large telescopes to search for activity in the dark comets.}
\label{fig:152667_observability}
\end{figure*}

If the dark comets are producing low levels of dust  or other observable outgassing activity, it is feasible that these will be detectable with future observations. In its extended mission, the Japanese Aerospace Exploration Agency's (JAXA) Hayabusa2 spacecraft (Hayabusa2$\#$) is scheduled to rendezvous with the dark comet  1998 KY$_{26}$ \cite{Hirabayashi2021,Kikuchi2023}. Hayabusa2$\#$ is expected to detect dust number densities as low as $10^5$ m$^{-3}$ \cite{Senshu2017}. This corresponds to a dust production sensitivity of $4.4\times10^{12}$ dust particles s$^{-1}$ assuming (i) the velocity of the outgassing material is $\sim$350 m s$^{-1}$, and (ii) the spacecraft approaches within $\sim$100 m of the target --- although advanced operations may include impacting and landing \cite{Kikuchi2023}.  The  H$_2$O production rate of 1998 KY$_{26}$ inferred from the nongravitational acceleration  is of order $10^{21}$ particles s$^{-1}$ (Figure \ref{fig:mass_loss_rate}). Therefore, Hayabusa2$\#$ should detect dust if the dust-to-gas ratio is larger than $>10^{-9}$. This mission will also characterize the surface and ambient environment of 1998 KY$_{26}$ \cite{Kameda2015,Watanabe2017,Arai2017,Okada2017,Takita2017,Mizuno2017,Senshu2017,Yamada2017,Kameda2017,Iwata2017,Suzuki2018,Tatsumi2019}. 

Moreover, these new detections of dark comets provide ample opportunity for follow-up observations. In Figure \ref{fig:152667_observability}, we show the timeline of observable apparitions (when the visual magnitude is brighter than 25, heliocentric distance is less than 1.2 au, and solar elongation angle is less than $60^\circ$) for all of the dark comets. Targeted observations that (1) identify outgassing/dust activity via deep spectroscopy/imaging, (2) further constrain the accelerations with astrometric measurements, and (3) obtain surface reflectance spectra to identify hydrated silicates or water ice driving the acceleration would be informative.

These lines of evidence support the hypothesis that the outer dark comets are a distinct  population to the inner dark comets. The inner and outer dark comets may represent different evolutionary stages of hydrated planetesimals from the main asteroid belt and/or the JFC region \cite{Taylor2024}.   It is believed that comets supplied a fraction of the Earth's oceans \cite{Marty2016} based on consistent and elevated Deuterium to Hydrogen (D/H) levels compared to the Earth's oceans in comets 103P/Hartley 2 \cite{Hartogh2011} and 67P/Churyumov-Gerasimenko \cite{Altwegg2015}. Moreover, organics have been identified on a variety of comets  in situ such as 81P/Wild 2  \cite{Sandford2006,Elsila2009} and 67P/Churyumov-Gerasimenko \cite{Altwegg2019}. It is possible that the dark comets are a previously unidentified  subclass of near-Earth comets representative of previous generations of small bodies that  delivered building blocks of life to the early Earth from the outer solar system. Regardless of these objects' specific role in volatile delivery, their discoveries point to a heretofore unrecognized population of weakly active objects in the inner solar system. This is similar to the proposition by \cite{sonnett2011_cfhtmbcs} that a large number of main-belt asteroids may exhibit extremely weak activity that has gone undetected to date. This would  indicate that the present-day inventory of active small bodies could be much larger than is currently understood.

To summarize our findings, we identified seven new dark comets (doubling the population) which provide evidence for  two  populations based on their orbits and sizes. Outer dark comets may be end members of a continuum in activity level, rather than something particularly distinct from other comets. On  the other hand, the inner dark comets could signify a new population  that could shed  light on important processes not currently understood.

\section*{Materials and Methods}

\subsection*{Determination of Nongravitational Accelerations}

Outgassing can significantly change the orbits of comets by causing a recoil effect which induces nongravitational accelerations. The orbits are typically fit with a parametric nongravitational model \cite{marsden1973} where the accelerations are defined as,
\begin{equation}\label{eq:forces}
\mathbf a_{\rm NG} = \left( A_1 \hat{\mathbf r} + A_2 \hat{\mathbf t} + A_3 \hat{\mathbf n}\right) \, g(r)\,.
\end{equation}
In Equation \ref{eq:forces}, the unit vectors $\hat{\mathbf r}$, $\hat{\mathbf t}$, and $\hat{\mathbf n}$ correspond to the radial, transverse, and out-of-plane directions respectively. The function $g(r)$ is some parametric function that depends on the heliocentric distance $r$ and is generally the empirical heliocentric distance dependence of H$_2$O activity. In this paper, we use $g(r)=(1 \text{ au}/r)^2$. However, the significance of the detections does not vary significantly by changing the $g(r)$ power law index. $A_1$, $A_2$, and $A_3$ are the magnitudes of the radial, transverse and out-of-plane components of the acceleration for $r=1$ au. 

\begin{figure}
\centering
\includegraphics[width=1.0\linewidth]{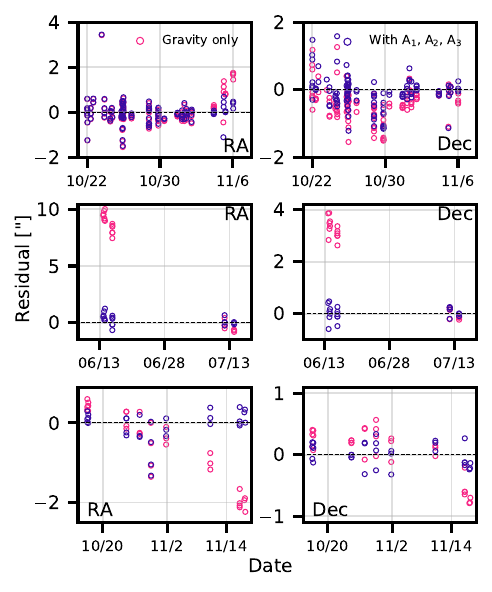}
\caption{The signed residuals of the astrometric fits  for 2012 UR$_{158}$ for sky position RA (left) and Dec (right). The points indicate the residuals for fits with only gravity (red circles) or with nongravitational accelerations (blue circles). The three sets of panels show different dates, corresponding to apparitions in 2012, 2016, and 2022 respectively.}
\label{fig:astrometric_residuals}
\end{figure}

In order to estimate nongravitational accelerations, the astrometric data over time is fit to model orbits with nongravitational accelerations by a least-squares method described in \cite{Farnocchia2015aste.book}. Details of this analysis for 2003 RM were presented in \cite{Farnocchia2022}. This estimation process  results in formal best-fit nongravitational acceleration components and corresponding uncertainties and statistical significances. As an example, in Figure \ref{fig:astrometric_residuals} we show the astrometric residuals in RA and Dec, for the object 2012 UR$_{158}$ during the course of three of its apparitions. These are shown with both the best-fit gravity-only orbit and the orbit with nongravitational accelerations.  The $\chi^2$ value for the fit of each object for both cases are shown in Table \ref{table:accelerations}. It is apparent that the residuals are significantly reduced upon incorporation of nongravitational accelerations.

In Table \ref{table:accelerations}, we show all of the dark comets' best-fitting nongravitational acceleration, uncertainties, and significance for all three acceleration components. The entire astrometric data sets are publicly available\footnote{\url{https://minorplanetcenter.net/iau/ECS/MPCAT-OBS/MPCAT-OBS.html} andd \url{https://ssd.jpl.nasa.gov/sb/radar.html}}. Optical astrometry was debiased to correct for star catalog biases \cite{Eggl2020}, weighted according to the Veres et al. (2017) scheme \cite{Veres2017}, and outliers rejected based on the Carpino et al. (2003) algorithm \cite{Carpino2003}.

\begin{table*}
\caption{Upper limit on the mass of dust that is compatible with the observations, for each image analyzed. Columns correspond to the JPL Small-Body Database identification number, provisional designation, date of observations, instrument, filter, exposure time, heliocentric distance, geocentric distance, true anomaly, full-exposure zeropoint magnitude, object magnitude, outer profile dust magnitude  and mass limit, inner profile dust magnitude and mass limit, and $3-\sigma$ surface brightness detection limit.} 
\label{table:dustMassDetail}
 \begin{tabular}{cccccccccccccccc}
 Number & Name & Date & Instrument & Filter & Exp.time & r & $\Delta$ & $\nu$ & ZP & Mag & Mag & M & Mag & M & $\Sigma_{\rm lim}$ \\
    &   & [UT]    &            &  & [s]      &[au]& [au]    & [deg] &   &Object & Dust in & [kg] & Dust out &[kg] & mag~arcsec$^{-2}$ \\
(139359) &\textbf{\textit{2001 ME$_{1}$}} & 2015-Mar-30 03:23 & DECam &$z'$ & 66.0 & 3.234 & 2.249 &156.6 & 30.0 & 21.2 & 22.3 & 6.2$\times 10^{3}$ & 21.4 & 1.5$\times 10^{4}$ & 22.0 \\
(139359) &\textbf{\textit{2001 ME$_{1}$}} & 2015-May-18 23:18 & DECam & $r'$ & 30.0 & 3.569 & 3.032 & 160.1 & 29.4 & 22.5 & 23.2 & 6.2$\times 10^{3}$ & 22.4 & 1.3$\times 10^{4}$ & 24.2\\
(139359) &\textbf{\textit{2001 ME$_{1}$}} & 2016-Apr-09 04:58 & DECam & $r'$ & 69.0 & 4.803 & 3.808 & 174.5 & 30.1 & 23.0 & 23.7 & 1.0$\times 10^{4}$ & 23.0 & 2.1$\times 10^{4}$ & 24.9 \\
(139359) &\textbf{\textit{2001 ME$_{1}$}} & 2022-May-03 08:16 & DECam & $g'$ & 50.0 & 3.558 & 2.556 & 199.5 & 28.7 & 21.6 & 22.6 & 7.6$\times 10^{3}$ & 22.1 & 1.1$\times 10^{4}$ & 24.6 \\
(139359) &\textbf{\textit{2001 ME$_{1}$}} & 2022-May-03 08:25 & DECam & $i'$ & 50.0 & 3.557 & 2.556 & 199.5 & 29.7 & 21.6 & 22.7 & 6.5$\times 10^{3}$ & 21.9 & 1.4$\times 10^{4}$ & 23.8 \\
(139359) &\textbf{\textit{2001 ME$_{1}$}} & 2022-May-12 07:49 & DECam & $r'$ & 50.0 & 3.501 & 2.497 & 200.1 & 29.9 & 21.5 & 22.5 & 7.5$\times 10^{3}$ & 22.7 & 6.1$\times 10^{3}$ &24.4 \\
(139359) &\textbf{\textit{2001 ME$_{1}$}} & 2022-May-12 08:22 & DECam & $i'$ & 50.0 & 3.500 & 2.497 & 200.1 & 30.1 & 21.5 & 22.9 & 5.0$\times 10^{3}$ & 22.1 & 1.1$\times 10^{4}$ & 23.7 \\
(139359) &\textbf{\textit{2001 ME$_{1}$}} & 2022-Jun-08 02:15 & DECam & $g'$ & 15.0 & 3.321 & 2.474 & 201.9 & 28.0 & 21.8 & 22.2 & 9.0$\times 10^{3}$ & 21.6 & 1.5$\times 10^{4}$ & 24.1 \\
(139359) &\textbf{\textit{2001 ME$_{1}$}} & 2022-Jun-08 02:16 & DECam & $g'$ & 15.0 & 3.321 & 2.474 & 201.9 & 27.9 & 21.8 & 22.1 & 9.4$\times 10^{3}$ & 21.4 & 1.8$\times 10^{4}$ & 24.1 \\
(139359) &\textbf{\textit{2001 ME$_{1}$}} & 2022-Jun-08 02:17 & DECam & $r'$& 15.0 & 3.321 & 2.474 & 201.9 & 28.7 & 21.8 & 22.6 & 6.1$\times 10^{3}$ & 22.0 & 1.0$\times 10^{4}$ & 23.7 \\
(139359) &\textbf{\textit{2001 ME$_{1}$}} & 2022-Jun-08 02:18 & DECam & $i'$ & 15.0 & 3.321 & 2.474 & 201.9 & 28.9 & 21.8 & 23.0 & 4.3$\times 10^{3}$ & 21.8 & 1.3$\times 10^{4}$ & 23.3 \\
(139359) &\textbf{\textit{2001 ME$_{1}$}} & 2022-Jun-08 02:19 & DECam & $i'$ & 15.0 & 3.321 & 2.474 & 201.9 & 28.8 & 21.8 & 21.9 & 1.1$\times 10^{4}$ & 21.3 & 1.9$\times 10^{4}$ & 23.3 \\
(139359) &\textbf{\textit{2001 ME$_{1}$}} & 2024-Mar-14 04:35 & FORS2 & Free & 1500.0 & 4.047 & 3.128 & 165.4 & 37.7 & 22.5 & 24.8 & 2.1$\times 10^{3}$ & 25.5 & 9.5$\times 10^{2}$ & 26.8 \\
(152667) &\textbf{\textit{1998 FR$_{11}$}} & 2016-Jan-15 05:54 & DECam & $r'$ & 43.0 & 4.104 & 3.169 & 201.6 & 29.3 & 22.4 & 23.1 & 9.1$\times 10^{3}$ & 22.6 & 1.5$\times 10^{4}$ & 24.7 \\
(152667) &\textbf{\textit{1998 FR$_{11}$}} & 2016-Jan-15 05:55 & DECam &$r'$ & 79.0 & 4.104 & 3.169 & 201.6 & 29.0 & 22.4 & 22.9 & 1.1$\times 10^{4}$ & 22.3 & 2.0$\times 10^{4}$ & 25.2 \\
(523599)& \textbf{2003~RM} & 2013-Aug-06 08:43 & DECam & $r'$ & 30.0 & 1.201 & 0.321&  118.4& 28.3& 19.4& 20.7&
   $7.9\times 10^1$& 21.46&  $3.8\times 10^1$& 23.4 \\
(523599)& \textbf{2003~RM} & 2018-Oct-13 03:29 & DECam & $r'$ &112.0 & 1.649 & 0.671&  161.7& 29.9& 20.5& 22.3&
   $1.4\times 10^2$& 22.11& $1.7\times 10^2$& 24.4 \\
 \end{tabular}
\end{table*}

\subsection*{Search for Dust Detection and Activity in Image Data}

We performed a search for detections and associated activity of the dark comets in archival and targeted imaging observations of the seven new objects. For our archival search, we limited our search to data obtained by the wide field Dark Energy Camera (DECam) on the  4~m Victor M.\ Blanco telescope in Chile and MegaCam on the 3.6~m Canada-France-Hawaii Telescope in Hawaii. These are the only two wide-field imagers on large-aperture telescopes that routinely perform surveys, and therefore provide the deepest and most comprehensive coverage of the sky of all currently operating facilities.   
Specifically, we used the Canadian Astronomy Data Centre's online Solar System Object Image Search (SSOIS) tool \cite{Gwyn2012} to search for archival DECam and MegaCam images containing serendipitous observations of these dark comets.

We compiled the data and searched for each object in each potential image. We made individual cutouts of the images centered on the objects and then subtracted the mean value of the background  from the image in order to search for the objects and detectable activity.

Additionally, targeted observations of 2001~ME$_1$ were obtained using the FOcal Reducer and low dispersion Spectrograph (FORS2) on the ESO Very Large Telescope at Paranal. The images were processed (bias subtraction and flat-field), and stacked, totalling 1500s exposure time. The stacked image and the associated radial profiles of the companion star and objects  are shown in Figure~\ref{fig:images_and_profiles}.

We searched by eye for any activity in all of the images described above. Careful visual evaluation of each image revealed no visible coma or extended emission such as tails or trails.  To further constrain the level of activity present in each image, the photometric profile of the object was compared to that of background stars, scaled in flux, and accounting for possible trailing of the object or the star. Examples of such profiles are displayed in Figure~\ref{fig:images_and_profiles}. This method is powerful enough to reveal barely resolved coma that would otherwise escape visual inspection. No profile showed any significant excess.

These profiles were used to quantify the amount of dust that could ``hide" in the profile, using two methods: (i) the dispersion on the average photometric profile points in the inner region (less than 1$''$ from the peak) is used as the error on the profile. Dust contribution smaller than that error would go unnoticed. (ii) The dispersion of the flux in the outer part of the profile (between 1.5$''$ and 2$''$) is used as the level below which an extended coma would go unnoticed. These measurements were performed on each image where the object is present with a signal-to-noise ratio of at least 3. For both methods, the flux in the studied region was converted into an absolute flux using the integrated flux of the object and its expected magnitude as calibration. The resulting photometric zero points were compared across images as a sanity check. To estimate the quantity of dust this corresponds to, we assumed that all the flux was caused by particles with a radius $a_d=1\,\mu$m, an albedo $p=0.02$ (typical value from Table~\ref{table:physical_properties}), and a density $\rho =$1\,000\,kg~m$^{-3}$ (the value for cometary grains measured by Rosetta \cite{Fulle2015}). Using the helio- and geocentric distance of the object ($r_h$ and $\Delta$, obtained from JPL's Horizon ephemerides service for the epoch of each observation), and a solar magnitude $M_\odot = -26.71$ (corresponding to a $V$ filter), the flux of a single grain is given by:
\begin{equation}\label{eq:fgrain}
    f_{\rm grain} = 10^{-0.4(M_\odot - ZP)} \frac{a_d^2 p}{r_h^2\Delta^2}\,.
\end{equation}
In Equation \ref{eq:fgrain}, $r_d$ is expressed in the same units as $r_h$ and $\Delta$, and $ZP$ is the full-exposure zeropoint magnitude.
The upper limits for the flux of the dust are converted to number of grains by dividing by $f_{\rm grain}$, and to mass by multiplying by the mass of a grain $m_{\rm grain} = 4\pi a_d^3 \rho/3$. There are large uncertainties on $a_d$ (and its distribution) and $\rho$, and the resulting values strongly depend on the size of the region considered (radii of the various aperture). Therefore, the results must be considered only as order-of-magnitude estimates of the upper limit of dust. Table~\ref{table:dustMassSummary} lists the most constraining upper limit for each object where estimates were possible. Table \ref{table:dustMassDetail} displays the values for each image, where we also list the $3-\sigma$ surface brightness detection limit ($\Sigma_{\rm lim}$) for each image (i.e., the maximum brightness for any tail or trail that could be present without being detectable) for reference.

At the heliocentric distances considered, the gas ejection velocity is of the order of 300--400~m\,s$^{-1}$. Assuming that the gas efficiently drags the grains, the grains will remain in the considered aperture for $10^3$-$10^4$ s, resulting in limits on the dust production rates of about 10~kg~s$^{-1}$.

To reach more constraining limits, the data obtained on each object were combined to produce deeper profiles, which were analyzed as above. The resulting upper limits on the dust (also listed in Table~\ref{table:dustMassSummary}) are an order of magnitude lower than those of the individual profile: of the order of $\sim10$, $0.1$ and $0.1$~kg~s$^{-1}$ at heliocentric distances of $\sim3.2-4.8$, $\sim4.1$ and $\sim1.2-1.6$ au  for  1998~FR$_{11}$, 2001~ME$_{1}$, and 2003~RM respectively.

\begin{figure}
\centering
\includegraphics[width=1.\linewidth]{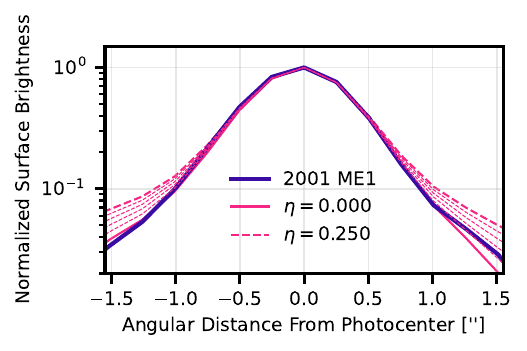}
\caption{Comparison of normalized surface brightness profiles of seeing-convolved model nuclei with varying levels of coma ranging from $\eta=0.00$ to $\eta=0.05$ in $\eta=0.05$ intervals (solid, dashed, and dotted pink lines) to that of 2001 ME$_{1}$ (solid purple line).}
\label{fig:coma_model}
\end{figure}

Finally, given the greater depth of our targeted VLT observations of 2001 ME$_{1}$, we perform an analysis in which we compared the surface brightness profile of that object to those of a template star with varying amounts of added coma. This provides an alternate method of determining the limit of the amount of excess dust that could be present.  This analysis follows the method of \cite{Luu1992} (also see \cite{hsieh2005_phaethon} and \cite{lilly2021_centaurs}).  In this analysis, we create synthetic active objects by (i) injecting varying amounts of additional dust coma to model point sources, (ii) convolving these models with a stellar PSF template created from the same VLT data as our target observation, and (iii) comparing the linear surface brightness profiles (measured perpendicular to the direction of 2001 ME$_{1}$'s non-sidereal motion to mitigate trailing effects) of those synthetic objects to that of our target object. This analysis provides upper limits on the amount of dust that could plausibly go undetected.  Coma levels were parameterized by $\eta=C_c/C_n$, where $C_c$ and $C_n$ are the scattering cross-sections of the coma and nucleus, respectively, and the reference photometry aperture radius used was $\Phi=6.25''$ (25 pixels).   Specifically, we tested coma levels of $\eta=0.00-0.25$ in $\eta=0.05$ increments.

Model profiles are overlaid on the target profile shown in Figure~\ref{fig:coma_model}. This demonstrates that deviations in the target profile from the nominal inactive stellar profile fall approximately within an envelope bounded by the $\eta=0.05$ profile, and therefore adopt this as our limiting coma parameter, $\eta_{\rm lim}$.  An upper limit dust production rate, $d M_d/dt$, can then be estimated from $\eta_{\rm lim}$ using
\begin{equation}\label{eq:dmddt}
    \frac{d M_{\rm d}}{dt} = \, \big( \, 1.1\times10^{-3}\,\big) {\frac{\pi\,\rho\,a_d\, \eta_{\rm lim}\,r_n^2}{\Phi\, r_h^{0.5}\, \Delta}}\,.
\end{equation}
\cite{Luu1992}. In Equation \ref{eq:dmddt}, $r_n$ is the object's effective radius, $\Phi$ is in arcseconds, and $r_h$ and $\Delta$ are in astronomical units.  Using values of $r_n\sim5\times10^3$ from Table~\ref{table:physical_properties},  $\rho=1$~g~cm$^{-3}$ and $a_d=1\times10^{-6}$~m from above, we estimate the visible upper limit mass loss rate from 2001 ME$_{1}$ from these observations to be $d M_d/dt\sim0.1$~kg~s$^{-1}$.

We note that deeper targeted observations in the future, particularly closer to perihelion, may reveal faint dust emission in one or more of these objects.  That said, we do note that such observations of 2003 RM totalling 3280~s in exposure time in 2018 when it was at $r_h=1.2$~au and $\Delta=0.3$~au still did not reveal any activity \cite{Farnocchia2022}, indicating that at least this object's activity, if present, genuinely appears to be  extraordinarily faint.

\subsection*{Search for Activity using Hyperstacking}

In order to perform a deeper search for activity using archival data, we use a novel enhanced approach to shifting and adding images (also referred to as image stacking) that we refer to as hyperstacking.  In this approach, rather than only adding images of a single object obtained on a single night, or even images of a single object obtained over multiple nights, we add images of multiple objects over multiple nights (e.g., all available data for all dark comets known to date) in order to maximize our sensitivity to low average activity levels that may be undetectable for individual objects (e.g., as has been suggested for main belt asteroids \cite{sonnett2011_cfhtmbcs}).  This approach draws inspiration from analyses using stacked images of sets of galaxies to enable the measurement of their average physical properties, despite having insufficient signal-to-noise to achieve meaningful measurements for individual galaxies \cite{zenteno2016_sptgalaxies,wang2016_stackinggalaxies}.

At this time, our analysis primarily aims to detect directional activity features, namely dust tails or trails, rather than spherical coma. This is due to complicated considerations associated with constructing meaningful average stellar reference profiles from observations obtained across a wide range of observational circumstances.  For active comets, tails and trails are commonly approximately aligned with the projections of either the antisolar vector (i.e., the direction opposite from the Sun) or the negative heliocentric velocity vector (i.e., aligned with the object's orbital plane). Roughly speaking, antisolar tails generally consist of small, short-lived ejected particles while orbit plane--aligned dust trails typically consist of larger, longer-lived ejected particles \cite{sykes1992_cometdusttrails}.  We can therefore use this property to combine images of different objects obtained at different times at disparate observing geometries in a way that may enable detection of extremely low-brightness activity features that are undetectable in individual images.

As a demonstration of this technique, we experiment with applying it to our set of dark comets.  We first identify images from a particular data set in which our targets are clearly detected, rotate them to place the expected direction of activity at a common orientation (in our case, directly to the right), and add the rotated images.  For completeness, we perform this analysis by aligning the images to both the antisolar and orbit plane directions (position angles obtained from JPL's Horizons online ephemeris tool \cite{giorgini1996_horizons}). As mentioned above, ejected dust aligned with the orbit plane should be longer-lived (assuming that dust particles large enough to be captured into dust trails are actually produced) and therefore more likely to be detected. Currently, we make no effort to match individual images other than rotating them into alignment and combing data obtained from the same telescope. However, additional matching methods (e.g., by normalizing source brightness in individual images or matching physical spatial scales by pixel resampling) may be included in future applications if they are shown to produce improved results.

We show our results using DECam data identified and retrieved using the SSOIS tool in Figures~\ref{fig:hyperstack_decam_images} and \ref{fig:hyperstack_decam_profiles}.  These data are specifically comprised of $g'$-, $r'$-, $i'$-, and $z'$-band data for four objects:  two images of 1998 FR$_{11}$ totalling 122~s of exposure time, nine images of 2001 ME$_{1}$ totalling 654~s of exposure time, two images of 2003 RM totalling 142~s of exposure time, and one image of 2012 UR$_{158}$ with an exposure time of 30~s, giving a total of 948 s of combined exposure time.
Only data in which objects were clearly visible by eye and not immediately adjacent to background sources are used.  Figure \ref{fig:hyperstack_decam_images}  shows median-combined data for our target objects as well as field stars of similar brightnesses from the same data (one field star per image) aligned on the antisolar direction and orbit plane direction.  
No evidence of directional activity features (antisolar or orbit plane directions) are found in either set of hyperstacks as demonstrated by azimuthal net surface brightness plots created for each hyperstack, also shown in Figure~\ref{fig:hyperstack_decam_profiles}. 

\begin{figure}
\centering
\includegraphics[width=1.\linewidth]{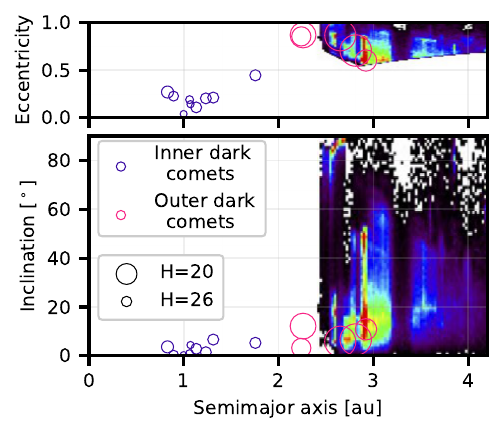}
\caption{The percentage chance that the object came from the JFC region based on the NEOMOD model \cite{Nesvorny2023}. The percentage chance is shown in the background color, where warmer colors indicate higher probabilities. The locations of the dark comets are indicated, as in Figure \ref{fig:tisserand}.}
\label{fig:JFCProb}
\end{figure}

The small number of relevant targets known at the moment and their limited observability (and therefore limited availability of serendipitous observations) limit the effectiveness of the hyperstacking approach for this particular work.  However, as more dark comets are discovered and additional data become available, this technique may prove more fruitful in detecting low-level activity from these objects.  For larger data sets, it will also become feasible to subdivide data into bins (e.g., by object size, rotation rate, or heliocentric distance) that may be relevant to activity detectability, improving focus. In the meantime, we suggest that searches for low average activity levels using this technique from myriad other populations that may contain mass-losing objects (e.g., fast rotators, asteroid families known to contain active asteroids, primitive-type near-Earth objects \cite{hergenrother2011_fastrotators,hsieh2018_activeastfamilies,lauretta2019_bennuactivity}) could be worthwhile.

\subsection*{Probability of Originating in the JFC Region}

In this subsection, we discuss how we estimated the percentage chance that each dark comet originated in the JFC region. In Figure \ref{fig:JFCProb}, we show the probabilities in semimajor axis, eccentricity, and inclination space with all of the currently known dark comets locations indicated. The JFC-origin probabilities are obtained from the NEOMOD model \cite{Nesvorny2023}. This model simulates the dynamical evolution of a population of NEOs injected from a variety of source populations (such as resonances in the main belt or JFC region). However, NEOMOD does not include nongravitational accelerations in its calculations. In addition, the JFC population is based on prior work \cite{Nesvorny2017}, and the terrestrial planets are not accounted for as a part of the dynamical evolution of that specific source. As a result, the probabilities here may be biased, and future work is necessary to fully clarify this population. The percentages are summarized in Table \ref{table:physical_properties}. Two of the outer dark comets show 0\% probability of JFC origin in Table~\ref{table:physical_properties}, but in view of their placement on Figure~\ref{fig:JFCProb} and the above caveats, this is not definitive as to their origin.

\subsection*{Data Availability Statement}
Data and figure producing scripts are available at \url{https://github.com/DSeligman/DarkComets.git}. The JPL asteroid and comet orbit determination code, used to estimate nongravitational parameters, is proprietary. However, the main results of this analysis can be reproduced by using freely available software such as OrbFit (\url{http://adams.dm.unipi.it/orbfit/})  and FindOrb (\url{https://www.projectpluto.com/find_orb.htm}).

\acknow{We thank the two anonymous reviewers for insightful comments and constructive suggestions that strengthened the scientific content of this manuscript. We thank Qicheng Zhang, Larry Denneau, Nikole Lewis, Jennifer Bergner, John Noonan, Luke Dones, Bill Bottke, Jonathan Lunine, Linda Glaser, Samantha Trumbo, Hal Levison,  Peiyu Wu, Fei Dai, Robert Jedicke, David Nesvorn\'y, David Vokrouhlick\'y,  Ngoc Truong, Robert Jedicke, Melissa Brucker, Jason Wright, Martin Seligman, David Jewitt, Jane Luu, and Yury Aglyamov   for useful conversations. We also thank   Marina Brozovic and Jon Giorgini for reviewing radar astrometry of a couple dark-comet candidates.  This research used the facilities of the Canadian Astronomy Data Centre operated by the National Research Council of Canada with the support of the Canadian Space Agency.

D.Z.S. is supported by an NSF Astronomy and Astrophysics Postdoctoral Fellowship under award AST-2303553. This research award is partially funded by a generous gift of Charles Simonyi to the NSF Division of Astronomical Sciences.  The award is made in recognition of significant contributions to Rubin Observatory’s Legacy Survey of Space and Time. A.G.T. acknowledges support from the Fannie and John Hertz Foundation and the University of Michigan's Rackham Merit Fellowship Program. The work of D.F. and S.R.C. was conducted at the Jet Propulsion Laboratory, California Institute of Technology, under a contract with the National Aeronautics and Space Administration (80NM0018D0004). H.~H. acknowledges support from NASA grants NNX17AL01G and 80NSSC19K0869. A.D.F. acknowledges  funding from NASA through the NASA Hubble Fellowship grant HST-HF2-51530.001-Awarded by STScI. K.J.M. acknowledges support from NASA grant 80NSSC18K0853.
}

\showacknow{} 


\bibliography{pnas-sample}

\end{document}